\documentclass[12pt]{iopart}
\usepackage{epsfig}
\usepackage{graphicx}
\usepackage{iopams}

 \newcommand \pss {{\mbox {\boldmath $\psi$}}}
 
 \newcommand \xis {{\mbox {\boldmath $\xi$}}}

 \newcommand{\order}{{\cal O}}
 \newcommand{\R}{{\rm I\!R}}
 \newcommand{\be}{\begin{equation}}
 \newcommand{\ee}{\end{equation}}
 
 \newcommand{\sgn}{{\rm sgn}}
 \newcommand{\pprime}{{\prime\prime}}

\begin{document}

\title{Asymmetrically extremely dilute neural networks with non-trivial dynamics}

\author{J P L Hatchett and A C C Coolen}
\address{Department of Mathematics, King's College, University
 of London, The Strand, London WC2R 2LS, United Kingdom}

\begin{abstract}
We study graded response attractor neural networks with
asymmetrically extremely dilute interactions and  Langevin
dynamics.  We solve our model in the thermodynamic limit using
generating functional analysis, and find (in contrast to the
 binary neurons case) that even in statics one cannot eliminate the non-persistent order
parameters. The macroscopic dynamics is
driven by the (non-trivial) joint distribution of neurons and
fields, rather than just the (Gaussian) field distribution.
 We calculate phase transition lines and present simulation results in support of our theory.
\end{abstract}


\eads{\mailto{hatchett@mth.kcl.ac.uk}, \mailto{tcoolen@mth.kcl.ac.uk}}

\section{Introduction}

Hopfield-type models \cite{Hopfield} are recurrent networks of
binary neurons (or Ising spins) with specific types of
pair-interactions, designed to store and retrieve information in a
distributed way. The main interest in their properties relates to
the regime of operation close to saturation, where the number
of patterns stored $p$ scales with the number of bonds per neuron $c$
(i.e. $\alpha=p/c$ remains finite as $N\to\infty$). Over the years
they have been studied intensively using statistical mechanical
tools, in various connectivity versions. Rigorous analysis, based
on techniques borrowed from spin-glass theory, has been carried
out for Ising spin neural networks with full connectivity
\cite{AGS1,AGS2,Riegeretal,Horneretal} ($c,N\to\infty$, $c/N=1$),
and for asymmetric \cite{Derrida,Kree} and symmetric extreme
dilution \cite{WatkinSherr} ($c,N\to\infty$, $c/N\to 0$). The
approach of \cite{Derrida,Kree} also proved effective for solving
models with continuous neurons and discrete time dynamics
\cite{Mertens}. More recently, attention has turned to finite
connectivity attractor neural networks \cite{Wemmenhove}
($c=\order(1)$, $N\to\infty$).

In this paper we study the dynamics of asymmetrically extremely
dilute graded response attractor neural networks with Langevin
dynamics,  near saturation. Such systems violate detailed balance
in two ways, firstly due to the interaction asymmetry and secondly
by the presence of a non-linear gain function (required for firing
rates to saturate). Equilibrium statistical mechanics no longer
applies; the system will never be in equilibrium. Yet, one still
expects evolution to a
 stationary state, albeit one with microscopic
probability currents.
The stationary state of fully connected networks of graded response
neurons with continuous time dynamics \cite{Hopfield84} close to
saturation have been studied  at $T=0$, \cite{ Kuehn90, Kuehnetal,
  KuehnBos93} together with dynamics at arbitrary $T$ but away from saturation
\cite{Coolen}. In the present case of asymmetric extreme dilution
one might expect the ideas of \cite{Derrida,Kree} to apply, and
anticipate the familiar simplification resulting from Gaussian
distributed fields. This turns out to be wrong. In
\cite{Derrida,Kree} and \cite{Mertens} the spins were
(stochastically) aligned to local fields in parallel, at discrete
intervals. Spins at time $t+1$ did not depend on spins at time $t$
other than via the fields. Due to the asymmetric dilution, the
field distribution was Gaussian  on finite time-scales
(characterized by two scalars), leading directly to simple
iterative laws for one or two scalar order parameters. In
contrast, although here the fields still have a Gaussian
distribution, the Langevin dynamics makes our model qualitatively
different: spin updates now depend strongly on present values, in
addition to the fields, and we will therefore need the
(non-Gaussian) joint spin-field distribution.

In this paper we use the generating functional analysis method of
\cite{Dominicis}, which has a strong record in the area of
asymmetric disordered spin systems (e.g.
\cite{Crisanti1,Crisanti2,Rieger,Duringetal}),
 following closely the exposition in \cite{Coolen} which allows us to be compact.
As always, in the infinite size limit one finds an effective single
spin equation, which in the special case of
complete asymmetry will have no retarded self-interaction. This
induces simplifications, resulting here in explicit
self-consistent laws for order parameter functions. The
 further ansatz of time translation invariance leads to a tractable
stationary state problem, from which, however, non-persistent
order parameters cannot be fully removed. In addition to solving
the full (exact) order parameter equations numerically,  we also
investigate an interpolative approximation that does depend on
only a few scalar quantities and compare the predictions of both
the full and the approximate theory with numerical simulation
data.   The interpolative theory is exact for zero $\alpha$ and
also at zero $T$ for small $\alpha$ (where small depends on the
choice of  gain function). We investigate the phase diagram of our
model in the $(\alpha, T)$ plane, and examine the transition from
non-recall to recall (we show that there is no spin-glass phase).
Numerical simulations support our theoretical findings and
predictions.

\section{Model definitions}

Our model describes a network of $N$ continuous neurons $u_i \in
\R$ which evolve in time according the following Langevin equation
\begin{equation}
\frac{\rmd}{\rmd t} u_i(t) = \sum_{j\neq i} J_{ij} g[u_j(t)] -
u_i(t) + \eta_i(t),
\label{eq:microdyn}
\end{equation}
where $g[u]$ is an odd sigmoidal function, which saturates for
$u\to \pm\infty$ to $\pm 1$ (respectively), and with the standard
Gaussian white noise terms $\eta_i(t)$ with moments
\begin{equation}
\langle \eta_i(t) \rangle = 0, \qquad \qquad \langle \eta_i(t)
\eta_j(t') \rangle = 2T\delta_{ij} \delta(t - t').
\end{equation}
The non-negative parameter $T$ controls the level of noise, with
$T=0$ and $T=\infty$ corresponding to deterministic and fully
random dynamics, respectively. The process ({\ref{eq:microdyn}})
does not obey detailed balance (not even for symmetric $J_{ij}$),
and thus will never reach equilibrium. This rules out the
techniques of equilibrium statistical mechanics,  and we are
forced to solve the dynamics. In the fashion of attractor neural
networks we store $p$ (randomly drawn) patterns  $\xis^{\mu} =
(\xi_1^{\mu},\ldots,\xi_N^{\mu}) \in \{-1,1\}^N$ in this system,
with $\mu = 1,\ldots,p$, by choosing diluted Hebbian-type
interactions $\{J_{ij}\}$ as in \cite{Derrida}:
\begin{equation}
J_{ij} =  \frac{c_{ij}}{c} \sum_{\mu = 1}^p
\xi_i^{\mu}\xi_j^{\mu}~~~~~~~~c_{ij}\in\{0,1\}
\end{equation}
The (fixed)  $\{c_{ij}\}$ define the connectivity, and are drawn
at random according to
\begin{eqnarray}
i<j: &~~~&   P(c_{ij}) = \frac{c}{N}\delta_{c_{ij},1} +
(1-\frac{c}{N}) \delta_{c_{ij},0}\\ i>j: &~~~&
P(c_{ij})=r\delta_{c_{ij},c_{ji}}+(1-r)\left[\frac{c}{N}\delta_{c_{ij},1}
+ (1-\frac{c}{N}) \delta_{c_{ij},0}\right]
\end{eqnarray}
The parameter $c$ is seen to specify the average number of
connections per spin, and  $r$ controls the symmetry of our
architecture. We will choose extreme dilution, i.e. $\lim_{ N
\rightarrow \infty} c/N =\lim_{N\to\infty} c^{-1}= 0$, such as
$c=\log (N)$. The number of patterns $p$ to be stored is expected
to scale with $c$,  so we define $p=\alpha c$ with $\alpha>0$
finite.

For simplicity, and in line with previous analyses, we will make
the so-called `condensed ansatz' in solving the dynamics: we
assume that the system state has an $\Or(N^0)$ overlap only with a
single pattern, say $\mu = 1$. This situation is induced by
initial conditions: we take a randomly drawn $\mathbf{u}(0)$,
generated by
\begin{equation}
\label{initcond}
p(\mathbf{u}(0)) = \prod_i \left\{\frac{1}{2}[1 + m_0]
\delta(u_i(0) - \xi_i^1) + \frac{1}{2}[1 - m_0]
\delta(u_i(0)+\xi_i^1) \right\},
\end{equation}
with $m_0$ measuring the overlap between the initial state and the
first pattern, $m_0 =N^{-1} \sum_i \langle \xi_i^1  u_i(0)
\rangle$. The patterns $\mu > 1$ and the architecture variables
$\{c_{ij}\}$, are viewed as disorder. We assume that for $N
\rightarrow \infty$ the macroscopic behaviour of the system is
`self-averaging', i.e. only depends on the statistical properties
of the disorder rather than its microscopic realization. Averages
over the disorder will be written as $\overline{\cdots}$.

\section{Generating functional analysis}

To analyze the dynamics following \cite{Dominicis} we  discretize
time (in units of $\Delta$), write the probability density of
observing a microscopic  `path'
$\{\mathbf{u}(0),\ldots,\mathbf{u}(t_m)\}$  through phase space as
$P[\mathbf{u}(0),\ldots,\mathbf{u}(t_m)]$,  and introduce the
familiar generating function $Z_{\Delta}[\pss]$:
\begin{eqnarray}
\hspace*{-15mm}
 Z_{\Delta} [\pss] &=& \overline{ \langle \exp\{-\rmi \sum_{i\leq N}
 \sum_{t\leq t_m}
\Delta \psi_i(t) g[u_i(t)]\}\rangle}
\nonumber
\\
\hspace*{-15mm} & =&  \int\! \rmd\mathbf{u}(0) \ldots
\rmd\mathbf{u}(t_m) \overline{
P[\mathbf{u}(0),\ldots,\mathbf{u}(t_m)] \exp\{-\rmi \sum_i \sum_t
\Delta \psi_i(t) g[u_i(t)]\} }\label{eq:genfun1}
\end{eqnarray}
Later we will put $\Delta\to 0$\footnote{The non-trivial
technicalities related to the assumed commutation of the limits
$N\to\infty$ and $\Delta\to 0$ are familiar issues in applying
saddle-point arguments to path integrals, which we will not
discuss here.}. Note that (\ref{eq:genfun1}) differs from the
standard definition in the appearance of the non-linear function
$g[u]$ in the exponent. For full details on the procedure for
deriving from (\ref{eq:genfun1}), in the limit $N\to\infty$, an
effective single spin equation we refer to e.g. \cite{Coolen}. The
discretized version of (\ref{eq:microdyn}) describes a Markov
process, so the path probability density is a product of
individual transition probability densities. In the
It$\hat{\mbox{o}}$ convention we do not pick up a Jacobian term
from the discretization.  We add time-dependent external fields
$\theta_i(t)\xi_i^1$ to the deterministic forces in order to
define response functions later, and proceed in the standard
manner to a saddle point problem. Solution of the latter leads us
to a closed macroscopic theory in terms of the familiar dynamic
order parameters $ m(t)= \lim_{N\to\infty}N^{-1} \sum_i \xi_i^1
\overline{\langle
 g[u_i(t)]\rangle}$ (the recall overlap),
 $C(t,t^\prime) = \lim_{N\to\infty}N^{-1} \sum_i \overline{\langle g[u_i(t)] g[u_j(t^\prime)]
\rangle}$ (the single-site correlation function) and
 $G(t,t^\prime) = \lim_{N\to\infty}\frac{1}{N} \sum_i
\partial\overline{g[u_i(t)]}/\partial \xi_i^1\theta_i(t)$
(the single-site response function). The order parameters are to
be solved from the saddle-point equations
\begin{eqnarray}
 m(t) &=& \langle g[u(t)] \rangle_{\star}
\label{eq:mdef}
\\
 C(t,t^\prime) &=& \langle g[u(t)] g[u(t^\prime)] \rangle_{\star}
  \label{eq:cdef}
\\
 G(t,t^\prime) &=&
\partial \langle
g[u(t)]\rangle_{\star}/\partial \theta(t^\prime) \label{eq:gdef}
\end{eqnarray}
Here the measure $\langle \ldots \rangle_{\star}$ is defined by
the statistics of the following effective single spin process
(having set $\pss = 0$ and taken the limit $\Delta \rightarrow
0$):
\begin{equation}
\frac{d}{dt} u(t) = -u(t) + m(t) + \theta(t) + \alpha r
\int_{-\infty}^t\! \rmd t^\prime~ G(t,t^\prime) g[u(t^\prime)] +
\phi(t) \label{eq:sspin1}
\end{equation}
where $\phi(t)$ is a zero mean Gaussian process with covariance
\begin{equation}
\label{phinoise} \langle \phi(t) \phi(t^\prime) \rangle = 2T
\delta(t - t^\prime) + \alpha C(t,t^\prime).
\end{equation}
In the remainder of this paper we will consider $r=0$ only (i.e.
fully asymmetric dilution, with $c_{ij}$  independent of
$c_{ji}$), so that the retarded self-interaction in
(\ref{eq:sspin1}) vanishes. This also implies that there is no
longer a need to solve for the response function, as the
macroscopic laws (\ref{eq:mdef},\ref{eq:cdef}) now close already
for the order parameters $\{m(t),C(t,t^\prime)\}$.

\section{Solving the single spin equation for asymmetric dilution}

As our main interest will be in stationary state properties, we
choose initial conditions at $t_0=-\infty$. For $r=0$ we can
readily integrate the stochastic equation (\ref{eq:sspin1}),
giving
\begin{eqnarray}
u(t) &=& k(t) +  Z(t) \label{eq:usoln}
\\
k(t) &=& \int_{-\infty}^t\!\! dt^\prime ~e^{-(t-t^\prime)}
[m(t^\prime) + \theta(t^\prime)]~~~~~~~ Z(t) =
\int_{-\infty}^t\!\! dt^\prime~ e^{-(t-t^\prime)} \phi(t^\prime)
\label{eq:defk} \end{eqnarray}
 So $Z(t)$ is also a zero mean
Gaussian noise, now with covariance $\langle
Z(t)Z(t^\prime)\rangle=\Xi(t,t^\prime)$
\be
\Xi(t,t^\prime) =  T e^{-\mid t-t^\prime \mid}
  + \alpha \int_{-\infty}^t\!\! ds~ e^{-(t-s)} \int_{-\infty}^{t^\prime}\!\!
  ds^\prime~
  e^{-(t^\prime - s^\prime)} C(s,s^\prime)
  \label{defxi2}
\ee We may use (\ref{eq:usoln},\ref{eq:defk}) and the standard
abbreviation $Dz = (2\pi)^{-1/2} e^{-\frac{1}{2}z^2}dz$ for
Gaussian measures to write the closed equations
(\ref{eq:mdef},\ref{eq:cdef}) in the following explicit form:
\begin{eqnarray}
m(t) &=& \int\! Dx~ g[k(t) + x\sqrt{\Xi(t,t)} ]
 \label{eq:mselfcon}
 \\
 C(t,t^\prime) &=& \int\! Dx Dy~  g[k(t) + a_1(t,t^\prime)(a_2(t,t^\prime)x + a_3(t,t^\prime)y)]
 \nonumber
\\ && \hspace*{11mm} \times~ g[k(t) +
 a_1(t,t^\prime)(a_2(t,t^\prime)x + a_4(t,t^\prime)y)]
 \label{eq:Cselfcon}
\end{eqnarray}
in which (using the symmetry of $\Xi$):
\begin{eqnarray}
 a_1(t,t') &=& 1/\sqrt{\Xi(t,t) + \Xi(t^\prime,t^\prime) -2\Xi(t,t^\prime)}\nonumber \\
 a_2(t,t') &=& \sqrt{\Xi(t,t)\Xi(t^\prime,t^\prime) - \Xi(t,t^\prime)^2}\nonumber\\
 a_3(t,t') &=& \Xi(t,t) - \Xi(t,t^\prime)\nonumber\\
 a_4(t,t') &=& \Xi(t^\prime, t) -\Xi(t^\prime,t^\prime)\nonumber
\end{eqnarray}

Let us turn to time translation invariant solutions (TTI), where
$m(t) = m$, $\theta(t) = \theta$ and $C(t,t^\prime) =
C(t-t^\prime)$, and consequently $\Xi(t,t^\prime) = \Xi(t -
t^\prime)$. This will simplify our analysis, and experience with
asymmetric systems suggests that most, if not all, solutions will
asymptotically obey TTI. We now find that  equations
(\ref{eq:mselfcon},\ref{eq:Cselfcon}) simplify to
\begin{eqnarray}
 m &=& \int\! Dx~ g[m + \theta + x\sqrt{\Xi(0)} ]
\label{eq:mtti}
\\
C(\tau) &=&  \int\! Dx Dy~ g\left[m + \theta +  x
\sqrt{\case12[\Xi(0) + \Xi(\tau)]} + y
 \sqrt{\case12[\Xi(0) - \Xi(\tau)] }\right]
 \nonumber
 \\
 &&~~~~~~\times~ g\left[m + \theta + x \sqrt{\case12[\Xi(0) + \Xi(\tau)]} - y \sqrt{\case12[\Xi(0) - \Xi(\tau)] }\right]
 \label{eq:ctti}
\\
\Xi(\tau) &=& Te^{-\mid \tau \mid} + \frac{1}{2}\alpha
\int_{-\infty}^{\infty}\! du~ e^{-\mid u + \tau\mid}C(u)
\label{xitti}
\end{eqnarray}
We note that, upon repeated differentiation with respect to
$\tau$, one can also derive from this latter equation  the
following identity:
\be
\Xi^{\pprime}(\tau)=\Xi(\tau)-\alpha C(\tau)
\label{eq:differential_form}
\ee
 In detailed  balance models at this stage
one would  be able to transform away the
non-persistent  order parameters, and derive closed equations for
the persistent objects $m$, $q_0\equiv C(0)$ and $q\equiv
C(\infty)$. Here such a reduction is not possible. To appreciate
this we first rewrite (\ref{eq:ctti}) (by suitably transformations
of integration variables) as
\be
C(\tau)= \int\!Dx\left\{
\int\!Dy~g\left[m+\theta+x\sqrt{\Xi(\tau)}+y\sqrt{\Xi(0)-\Xi(\tau)}\right]\right\}^2
\label{eq:Ctti_alternative} \ee Using the relation
$\Xi(\infty)=\alpha q$ and the short-hand $\Xi(0)=\kappa$, our
expressions for the persistent quantities $\{m,q_0,q\}$ then take
the form
\begin{eqnarray}
m&=&
 \int\! Dx~ g[m + \theta + x\sqrt{\kappa} ],
~~~~~~ q_0=
 \int\!Dx~g^2[m+\theta+x\sqrt{\kappa}]
\label{eq:mq0_alternative}\\
 q&=&
 \int\!Dx\left\{
\int\!Dy~g\left[m+\theta+x\sqrt{\alpha q }+y\sqrt{\kappa-\alpha
q}\right]\right\}^2 \label{eq:q_alternative}
\end{eqnarray}
with
\be
\kappa= T + \alpha \int_{0}^{\infty}\! d\tau~ e^{- \tau }C(\tau)
\label{eq:kappa} \ee Unless $T=0$ (where one has solutions with
$q=q_0$ and $C(\tau)=q$ for all $\tau$, to which we will return
later), the term $\kappa$ can not be expressed in terms of
$\{m,q_0,q\}$, but requires knowledge of non-persistent
correlations $C(\tau)$. From this stage onwards we will take
$\theta=0$, in order to study different solution types and
transitions between them.

\section{Special solutions}

\subsection{Non-recall TTI solutions}

 Non-recall
solutions are those where $m=0$ (to be expected for large $T$),
and where in the TTI regime we are left only with a closed set of
equations for $C(\tau)$:
\begin{eqnarray}
 C(\tau) &=&  \int\! Dx Dy~
g\left[x \sqrt{\case12[\Xi(0) + \Xi(\tau)]} + y
 \sqrt{\case12[\Xi(0) - \Xi(\tau)] }\right]\nonumber
 \\
 && ~~~~~~\times~ g\left[x \sqrt{\case12[\Xi(0) + \Xi(\tau)]} - y \sqrt{\case12[\Xi(0) - \Xi(\tau)] }\right]
 \label{eq:ctti_mzero}
\\
\Xi(\tau) &=& Te^{-\mid \tau \mid} + \frac{1}{2}\alpha
\int_{-\infty}^{\infty}\! du~ e^{-\mid u + \tau\mid}C(u)
\label{xitti_mzero}
\end{eqnarray}
The simplest case is $T \rightarrow \infty$. Since $C(\tau)$
is bounded, we may expand our equations and find that the second
term in (\ref{xitti_mzero}) is negligible relative to the first.
This results in
\begin{equation}
\lim_{T\to\infty} C(\tau) = \frac{2}{\pi} \arcsin (e^{-\mid \tau
\mid}) \label{eq:chight}
\end{equation}
Clearly $\lim_{T\to\infty}C(0)=1$ and
$\lim_{T\to\infty}C(\infty)=0$. It is a trivial matter to show
that this is the only possible solution for $T\to\infty$.

In the  case where $g[u] = \sgn[u]$ we can push the
analysis of (\ref{eq:ctti_mzero},\ref{xitti_mzero}) further for
arbitrary $T$, as we may do the integrations in
(\ref{eq:ctti_mzero}) explicitly (in the recall phase this would
not have been possible). For  $g[u] = \sgn[u]$ we may write
\begin{eqnarray}
C(\tau) &=& \int\! DxDy~\sgn[x + \tan(\psi)y] \sgn[x-\tan(\psi)y]
\label{eq:C_sgn}\\
 \tan^2(\psi) &=& [\Xi(0) - \Xi(\tau)]/[\Xi(0) + \Xi(\tau)]
\end{eqnarray}
Upon writing $(x,y)$ in polar coordinates,  the integrations can
be done, and we find
\begin{equation}
\sin(\frac{1}{2}\pi C(\tau)) = \frac{Te^{-\mid \tau \mid} +
\frac{\alpha}{2} \int_{-\infty}^{\infty}\! du~ \rme^{-\mid u +
\tau \mid} C(u)}{T + \frac{\alpha}{2} \int_{-\infty}^{\infty}\!
du~ \rme^{-\mid u \mid} C(u)} \label{eq:sgn_case}
\end{equation}
For $T\to\infty$ we recover (\ref{eq:chight}). Putting
$\tau\to\infty$ in (\ref{eq:sgn_case}) gives
\begin{equation}
\sin(\frac{1}{2}\pi C(\infty)) = \frac{\alpha C(\infty)} {T +
\alpha\int_{0}^{\infty}\! du~ \rme^{- u} C(u)}
\label{eq:sgn_case2}
\end{equation}
The paramagnetic (P) state $C(\infty)=0$ always solves
(\ref{eq:sgn_case}), as expected. It also follows from
(\ref{eq:sgn_case}) that a continuous transition from a
paramagnetic to a spin-glass (SG) state $\{m=0,~C(\infty)> 0\}$
would occur (unless preceded by a recall transition) at
$\kappa_c=2\alpha/\pi$.

\subsection{Detailed balance type solutions at zero temperature}

Let inspect whether and where we have detailed balance type
macroscopic laws, where after elimination of transient parts one
finds `effective' equations with $C(\tau>0)=q$. Here the ansatz
$C(\tau>0)=q$ implies that
 $\Xi(\tau) = Te^{-\mid \tau \mid} + \alpha q$ and $\kappa=T+\alpha q$. One now finds that
 it gives proper solutions of our order parameter equations if and
 only if $T=0$, in which case we obtain (for $\theta=0$) $q_0=q$
 and the familiar looking
\be
m=
 \int\! Dx~ g[m  + x\sqrt{\alpha q} ]
~~~~~~~~~
 q=
 \int\!Dx~g^2[m+x\sqrt{\alpha q }]
 \label{eq:DBT}
\ee For $g[u]=\tanh[\beta u]$ these equations (\ref{eq:DBT}) are
identical to those of the {\em symmetrically} extremely diluted
attractor network with binary neurons \cite{WatkinSherr}.
 For $g[u]=\sgn[u]$ we immediately find the three possible
solutions P $\{m=0,q=0\}$, SG $\{m=0,q=1\}$ (both P and SG exist
for any $\alpha$) and R $\{m=m^\star,q=1\}$, where $m^\star$ is
the solution of
 \be
  m={\rm Erf}[m/\sqrt{2\alpha }]
 \ee
The retrieval solution bifurcates from the SG one at
$\alpha_c=2/\pi\approx 0.637$.

The constant correlation function
implies that the spins are microscopically frozen, thus there are no
probability currents at all in the system so detailed balance is
trivially restored. It is clear that such solutions always exist and
are stable at $\alpha =0$.

We next inspect the local stability of the above special $T=0$
solutions against perturbations away from $C(\tau>0)=q$ for $\alpha >
0$. We thus put
 $m \rightarrow m + \delta m$ and $\Xi(\tau) \rightarrow  q
+ \delta \Xi(\tau)$ (with $\delta m$ and $\delta \Xi(\tau)$
small) and find the eigenvalue problem
\begin{equation}
\label{bifzeroT} \delta \Xi(\tau) = K_1 \int_{-\infty}^{\infty}\!
\rmd u ~\delta \Xi(u) [\frac{K_2}{1-K_2} \rme^{-\mid u \mid} + \rme^{-\mid \tau -
u \mid}]
\end{equation}
where, with $y(x) = m + x\sqrt{\alpha q} $
\begin{eqnarray}
K_1 &=& \frac{1}{2}\alpha \int\! Dx~\Big\{ g^\prime[y] \Big\}^2
\\
K_2 &=& \alpha \int \!Dx~g[y]g^{\prime\prime}[y] +
\frac{\sqrt{\frac{\alpha}{q}}\int\!Dx~ g^\prime[y] \ldotp \int\!
  Dx~g[y]g^{\prime\prime}[y]}{1 - \int\!Dx~g^\prime[y]}
\end{eqnarray}
The only eigenfuction of (\ref{bifzeroT}) which is not trivial
 and bounded for $\tau \rightarrow \infty$
is given by
\begin{equation}
\delta \Xi(\tau) = \cos(\sqrt{2K_1-1}\tau) -
\frac{K_2}{1-K_2-2K_1}
\end{equation}
with $K_1\geq \frac{1}{2}$.  Since $\lim_{\alpha\to 0}K_1=0$, the
bifurcation condition becomes $K_1=\frac{1}{2}$, or
\be
1 =\alpha \int\! Dx~\Big\{ g^\prime[m+x\sqrt{\alpha q }] \Big\}^2
\label{eq:DBtrans} \ee
 The size of
the region along the $T=0$ axis where the solution (\ref{eq:DBT})
is stable is dependent on our choice of the gain function $g$. For
$g[u] = \sgn[u]$ we find that there is never a continuous
transition while for $g[u] = \tanh[\gamma u]$ there can be and the
solution (\ref{eq:DBT}) may be stable in a large part of the
recall region. To see more clearly why there are no continuous
transitions for $g[u] = \sgn[u]$ we note that in that case the
special solution is $q = q_0 = 1$, and our eigenvalue problem
becomes
\begin{equation}
\label{sgnspecsol} \alpha + \delta \Xi(\tau) = \frac{\alpha}{2}
\int_{-\infty}^\infty~\rmd u~ \rme^{-|u + \tau|}\int\!Dx~ {\rm
Erf}^2\left( \frac{m + x \sqrt{\alpha +
    \delta\Xi(u)}}{y\sqrt{\delta\Xi(0) - \delta\Xi(u)}} \right)
\end{equation}
For small perturbations the argument of the error function will be
large and we can use the expansion \cite{FFP}
\begin{equation}
{\rm Erf}(x) = 1 - \frac{1}{\sqrt{\pi}x}e^{-x^2} \sum_{n =
  0}^\infty\! \left(1 - \frac{1}{2x^2} + \frac{1\ldotp3}{(2x^2)^2} -
  \frac{1\ldotp3\ldotp5}{(2x^2)^3} + \ldots \right)
\end{equation}
to see that on the right-hand side of (\ref{sgnspecsol}) there is
no polynomial term in the perturbation, and hence no continuous
bifurcation.

\section{Non-equilibrium phase transitions}

\subsection{Continuous transitions without anomalous response}

Continuous transitions from P or SG states to recall states (R,
where $m>0$) or from P states to SG states, are found upon
expanding either (\ref{eq:mq0_alternative}) or
(\ref{eq:q_alternative}) (both with $\theta=0$) for small $m$ or
small $q$, respectively.  This is found to give the simple
bifurcation conditions
\begin{eqnarray}
{\rm P}\to {\rm R},~{\rm SG}\to {\rm R}:&~~~&  \int\! Dx
~g^\prime[x\sqrt{\kappa}]= 1 \label{mbif}\\ {\rm P}\to {\rm
SG}:&~~~&  \int\! Dx ~g^\prime[x\sqrt{\kappa}]= 1/\sqrt{\alpha}
 \label{qbif}
\end{eqnarray}
For $g[u]=\sgn[u]$ these conditions reduce to $\kappa_c=2/\pi$ and
$\kappa_c=2\alpha/\pi$, respectively. Here, since
$\kappa\in[T,T+\alpha q_0]$,
 we are sure that continuous transitions P$\to$R or SG$\to$R
 can occur at most for $T\in [2/\pi-\alpha,2/\pi]$, and that
 continuous transitions P$\to$SG
 can occur only when $T<2\alpha/\pi$.
To determine which instability away from the paramagnetic state
occurs first we need to inspect the dependence of $\kappa$, as
defined in (\ref{eq:kappa}) on $(\alpha,T)$.

\subsection{Transitions to states with anomalous response}

We next inspect the solution of our full TTI equations for large
times, upon separating non-persistent from persistent terms in the
functions $\Xi(\tau)$ from $C(\tau)$:
\be
C(\tau)=q+\tilde{C}(\tau),~~~~~~~~\Xi(\tau)=\alpha q
+\tilde{\Xi}(\tau) \ee Insertion into our equations for
$\Xi(\tau)$ and $C(\tau)$ gives
\begin{eqnarray}
\hspace*{-15mm} \tilde{\Xi}(\tau) &=& Te^{-\mid \tau \mid} +
\frac{1}{2}\alpha \int_{-\infty}^{\infty}\! du~ e^{-\mid u +
\tau\mid}\tilde{C}(u) \label{eq:Xitilde}
\\
\hspace*{-15mm} \tilde{C}(\tau)&=& \int\!Dx\left\{
\int\!Dy~g\left[m+x\sqrt{\alpha q+
\tilde{\Xi}(\tau)}+y\sqrt{\kappa-\alpha q -
\tilde{\Xi}(\tau)}\right]\right\}^2-q \label{eq:Ctilde}
\end{eqnarray}
For large times the non-persistent terms will be small, so we may
expand the right-hand side of (\ref{eq:Ctilde}) in powers of
$\tilde{\Xi}(\tau)$. First we work out the nonlinearity
$g[\ldots]$:
\begin{eqnarray}
\hspace*{-10mm} g[\ldots]&=&g\left[m+x\sqrt{\alpha
q}+y\sqrt{\kappa-\alpha q}\right]\nonumber
\\
\hspace*{-10mm} &&\hspace*{-5mm}
+\frac{1}{2}\tilde{\Xi}(\tau)\left[ \frac{x}{\sqrt{\alpha q}}
-\frac{y}{\sqrt{\kappa-\alpha q}}\right]
g^\prime\left[m+x\sqrt{\alpha q}+y\sqrt{\kappa-\alpha q}\right]
+\order(\tilde{\Xi}^2)
\end{eqnarray}
which gives, after partial integration over the Gaussian disorder
variables,
 \begin{eqnarray}
\tilde{C}(\tau)&=& \Lambda~
\tilde{\Xi}(\tau)+\order(\tilde{\Xi}^2(\tau)) \\
\Lambda&=&\int\!Dx\left\{ \int\!Dy~g\left[m+x\sqrt{\alpha
q}+y\sqrt{\kappa-\alpha q }\right]\right\} \nonumber
\\
&&\times \left\{\int\!Dy\left[ \frac{x}{\sqrt{\alpha q}}
-\frac{y}{\sqrt{\kappa-\alpha q}}\right]
g^\prime\left[m+x\sqrt{\alpha q}+y\sqrt{\kappa-\alpha
q}\right]\right\} \nonumber
\\
&=& \int\!Dx~\left\{ \int\!Dy~g^\prime\left[m+x\sqrt{\alpha
q}+y\sqrt{\kappa-\alpha q }\right]\right\}^2 \label{eq:Lambda}
\end{eqnarray}
For sufficiently large times we may therefore replace the duo
(\ref{eq:Xitilde},\ref{eq:Ctilde}) by the leading order
\be
\tilde{C}(\tau) =\Lambda\left\{ Te^{-\mid \tau \mid} +
\frac{1}{2}\alpha \int_{-\infty}^{\infty}\! du~ e^{-\mid u +
\tau\mid}\tilde{C}(u)\right\} +\ldots \label{eq:Ctilde_closed} \ee
This equation is solved asymptotically by $\tilde{C}(\tau)\sim
e^{-\gamma \tau}$, with exponent $\gamma=\sqrt{1-\alpha\Lambda}$. We
must conclude that a transition to a regime with anomalous
response (which in equilibrium disordered systems would correspond
to the AT line \cite{AT}) occurs when $\alpha\Lambda=1$, i.e. when
\be
\alpha \int\!Dx~\left\{ \int\!Dy~g^\prime\left[m+x\sqrt{\alpha
q}+y\sqrt{\kappa-\alpha q }\right]\right\}^2=1
\label{eq:anomalous} \ee We see that in the paramagnetic state
$m=q=0$ the condition (\ref{eq:anomalous}) coincides with that of
the  P$\to$SG transition  (\ref{qbif}), and that for $T=0$ it
coincides with the instability condition (\ref{eq:DBtrans}).
Again, for the special choice $g[u]=\sgn[u]$ all integrals can be
done analytically, and we find (\ref{eq:anomalous}) taking the
more explicit form
\be
\alpha e^{-m^2/(\kappa+\alpha q)}=\frac{1}{2}\pi
\sqrt{\kappa^2-\alpha^2 q^2} \label{eq:anomalous_for_sgn} \ee

\section{Bounds and approximations for the stationary state}

\subsection{Arbitrary non-linearities}

We established that it is not possible to obtain closed equations
for persistent order parameters only, due to the dependence of
(\ref{eq:kappa}) on the short-time part of $C(\tau)$. However, we
know that $C(\tau)$ decays from $C(0)=q_0$ to $C(\infty)=q$, so
that (\ref{eq:kappa}) can be written as
\begin{eqnarray}
\kappa&=&T+\alpha q+\alpha \Delta(q_0-q),~~~~~~~~\Delta\in[0,1]
\label{eq:Delta_definition}
 \end{eqnarray}
 with $\Delta\approx 1$ when the decay of $C(\tau)$ is much slower than $\exp[-\tau]$,
 and $\Delta$ smaller when the decay  is fast.
Thus we know that our phase diagram must interpolate between those
obtained for the extreme cases $\Delta\in\{0,1\}$, with these
cases  becoming identical both for $T\to 0$ at small $\alpha$ (where
 $q_0-q\to 0$ before the transition to non-flat $C(\tau)$)
and $\alpha\to 0$:
\begin{eqnarray}
\hspace*{-18mm} {\rm `Slow'~limit},~\Delta=1:~~~ && m = \int\!
Dx~g[m+x\sqrt{T+\alpha q_0}]\\
 \hspace*{-18mm}  && q_0=\int\!Dx~g^2[m+x\sqrt{T+\alpha q_0}]  \\
 \hspace*{-18mm}  && q= \int\!Dx\left\{ \int\!Dy~g[m+x\sqrt{\alpha
q}+y\sqrt{T+\alpha(q_0-q)}]\right\}^2
\\
\hspace*{-18mm} {\rm `Fast'~limit},~\Delta=0:~~
  &&  m= \int\! Dx~ g[m+x\sqrt{T+\alpha q} ]\\
 \hspace*{-18mm} && q_0=\int\!Dx~g^2[m+x\sqrt{T+\alpha q}]  \\
 \hspace*{-18mm} && q=  \int\!Dx\left\{ \int\!Dy~g[m+x\sqrt{\alpha q}+y\sqrt{T}]\right\}^2
\end{eqnarray}
Rather than analyze the extreme bounding cases
$\Delta\in\{0,1\}$, one could construct a rational interpolation
between $\Delta=0$ and $\Delta=1$ by using our knowledge of the
long-time behaviour of the non-persistent correlations as
established in the derivation of (\ref{eq:anomalous}), viz.
$\tilde{C}(\tau)\sim e^{- \tau\sqrt{1-\alpha\Lambda}}$ with $\Lambda$
as given in (\ref{eq:Lambda}), to define the simplest function
which satisfies both the long-time profile and the initial
conditions $\tilde{C}(0)=q_0-q$: $\tilde{C}(\tau)=(q_0-q)e^{-
\tau\sqrt{1-\alpha\Lambda}}$. This implies the following approximate
expression for $\Delta$:
\begin{eqnarray}
\Delta&=& \int_0^\infty\!d\tau~e^{- \tau[1+\sqrt{1-\alpha\Lambda}]}
=\frac{1}{1+\sqrt{1-\alpha\Lambda}}
 \end{eqnarray}
Equivalently: \begin{eqnarray}
 \kappa&=&T+\alpha q+
\frac{\alpha(q_0-q)}{1+\sqrt{1-\alpha\Lambda}}
\label{eq:kappa_interpolation}
\end{eqnarray}
 This then leads us to the following theory:
\begin{eqnarray}
\hspace*{-18mm} {\rm interpolation}:~~
 && m= \int\! Dx~ g\Big[m + x\sqrt{T+\alpha q+
\frac{\alpha(q_0-q)}{1+\sqrt{1-\alpha\Lambda}}} \Big]\\
\hspace*{-18mm}
 && q_0=\int\!Dx~g^2\Big[m+x\sqrt{T+\alpha q+
\frac{\alpha(q_0-q)}{1+\sqrt{1-\alpha\Lambda}}}\Big] \\
\hspace*{-18mm} && q=
 \int\!Dx\left\{
\int\!Dy~g\Big[m+x\sqrt{\alpha q }+y\sqrt{T+
\frac{\alpha(q_0-q)}{1+\sqrt{1-\alpha\Lambda}}}\Big]\right\}^2
\\
\hspace*{-18mm}&& \Lambda=\int\!Dx~\left\{
\int\!Dy~g^\prime\Big[m+x\sqrt{\alpha q}+y\sqrt{T+
\frac{\alpha(q_0-q)}{1+\sqrt{1-\alpha\Lambda}} }\Big]\right\}^2
\end{eqnarray}
Again we note that these equations are exact in either of the
limits $T\to 0$ and $\alpha\to 0$.

\subsection{Predictions for $g[u]=\sgn[u]$}

For the choice $g[u]=\sgn[u]$,
where Gaussian integrals can be done, we have $q_0=1$. Thus the slow
limit gives the P $\to$ F transition line $\kappa = \alpha + T =
2/\pi$. In the paramagnetic state $q = 0$ thus the fast limit gives a
P$ \to$ F transition for $\alpha <
1$ at $T = 2/\pi$ and a P $\to$ SG transition for $\alpha > 1$ at $T =
2\alpha /\pi$. In addition, the
above set of interpolating equations can be simplified to
\begin{eqnarray}
m&=&
 {\rm Erf}[m/\sqrt{2\kappa}],
~~~~~~~~
 q=
 \int\!Dx~{\rm Erf}^2[\frac{m+x\sqrt{\alpha q }}{\sqrt{2(\kappa-\alpha
 q)}}]
 \label{eq:sgn_mq}
\\
 \kappa
&=& T+\alpha\left\{ \frac{(\kappa^2-\alpha^2 q^2)^{\frac{1}{4}}
+q(\sqrt{\kappa^2-\alpha^2 q^2}-\frac{2\alpha}{\pi}
e^{-m^2/(\kappa+\alpha q)})^{\frac{1}{2}}}{(\kappa^2-\alpha^2
q^2)^{\frac{1}{4}}+(\sqrt{\kappa^2-\alpha^2
q^2}-\frac{2\alpha}{\pi} e^{-m^2/(\kappa+\alpha
q)})^{\frac{1}{2}}}\right\} \label{eq:sgn_kappa}
\end{eqnarray}

\begin{figure}[t]
\vspace*{7mm} \setlength{\unitlength}{0.1cm}
\begin{picture}(60,60)
\put(40,5){\includegraphics[height=6cm,width=7cm]{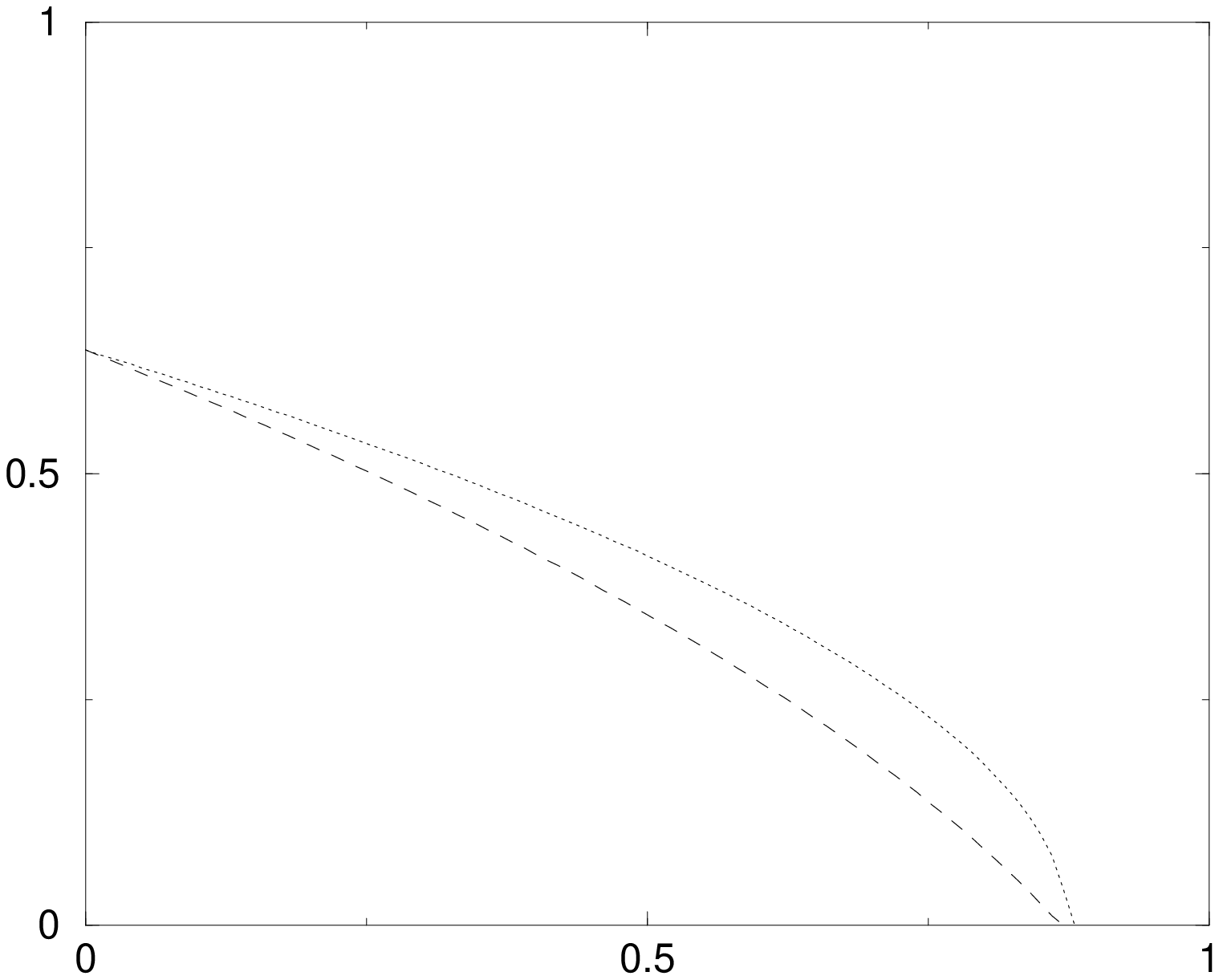}}
\put(76,0){$\alpha$}\put(35,35){$T$} \put(55,20){R}\put(75,40){P}
\end{picture}
\caption{\label{PHASED}
         Phase diagram for the asymmetrically dilute graded response
         model. The dashed line is the transition predicted by the
         interpolation approximation while the dotted line is that
         predicted by numerically solving for the correlation
         function. The interpolative method gives reasonable
         agreement
         but systematically underestimates the size of the recall
         region with the error increasing as a function of $\alpha$.}
\end{figure}

with the phase transitions occurring at:
\begin{eqnarray}
{\rm P}\to {\rm R},~{\rm SG}\to {\rm R}:&~~~&
\kappa_c=\frac{2}{\pi} \label{mbif_interp}
\\ {\rm P}\to {\rm SG}:&~~~& \kappa_c=\frac{2\alpha}{\pi}
 \label{qbif_interp}
 \\
{\rm Anomalous~Response}: &~~~& \sqrt{\kappa^2-\alpha^2
q^2}=\frac{2\alpha}{\pi} e^{-m^2/(\kappa+\alpha q)}
\label{eq:ar_interp}
\end{eqnarray}
In the paramagnetic state $m=q=0$, where the conditions
(\ref{qbif_interp},\ref{eq:ar_interp}) become identical, equation
(\ref{eq:sgn_kappa}) is seen to simplify significantly, and can in
fact  be solved for $\kappa$:
\begin{eqnarray}
\kappa &=&
\frac{T(1-\frac{2}{\pi})+\frac{1}{2}\alpha+\sqrt{T^2+\alpha
T(1-\frac{2}{\pi})+\frac{1}{4}\alpha^2}}{2(1-\frac{1}{\pi})}
\label{eq:sgn_kappa_P}
\end{eqnarray}
Given this expression we may in turn derive explicit expressions
for the our phase transition transition lines. One finds that the
transition P$\to$SG cannot occur, and that the P$\to$R transition
line reduces to
\be
{\rm P}\to {\rm R}:~~~~~
T_{c}(\alpha)=\sqrt{1-\alpha}-1+\frac{2}{\pi} \label{eq:phaseboundary}\ee This critical
temperature decreases with increasing $\alpha$ from its maximum
value $T_c(0)=2/\pi\approx 0.637$ down to $T_c(\alpha_c)=0$, with
 $\alpha_c=\frac{4}{\pi}(1-\frac{1}{\pi})\approx 0.868$.
Since our interpolation equations are exact at $T=0$ and at
$\alpha=0$, we can be sure that also the two critical values
$T_c(0)=2/\pi$ and $\alpha_c$ are exact.

 Thus, given that one
always enters a recall phase directly (without a spin-glass phase,
unlike symmetrically connected models), the only remaining
transition to be investigated is the one marking the possible
onset of anomalous response inside the retrieval phase. Combining
(\ref{eq:sgn_kappa}) with (\ref{eq:ar_interp}) shows that at this
transition $\kappa=T+\alpha$, so that we have to solve the
transition line from the following trio of coupled equations
\begin{eqnarray}
\hspace*{-10mm}
 {\rm Anomalous~Response}: &~~~& m={\rm
Erf}\Big[\frac{m}{\sqrt{2(T+\alpha)}}\Big]\label{eq:Anom_resp1}\\ &~~~& q=\int\!Dx~{\rm
Erf}^2\Big[\frac{m+x\sqrt{\alpha q}}{\sqrt{2[T+\alpha(1-q)]}}\Big] \label{eq:Anom_resp2}
\\
&~~~&
[T+\alpha(1-q)][T+\alpha(1+q)]=\frac{4\alpha^2}{\pi^2}e^{-\frac{2m^2}{T+\alpha(1+q)}}
 \label{eq:Anom_resp3}
\end{eqnarray}
Examining (\ref{eq:kappa}) we see that for an anomolous response we
require $q = 1$ which via (\ref{eq:Anom_resp2},\ref{eq:Anom_resp3}) implies
$T=\alpha=0$ is the only possibility. However, there one trivially has
$C(\tau) = 1, \forall \tau$ and there is no transient term in the
correlation function.

We plot the phase diagram Figure \ref{PHASED}, both using the approximation
(\ref{eq:phaseboundary}) and by solving for the whole correlation
function numerically. We see that the results are similar although
not identical, the interpolation method seems to
predicts a slightly smaller recall region.

\begin{figure}[t]
\vspace*{3mm} \setlength{\unitlength}{0.1cm}
\begin{picture}(60,60)
\put(10,5){\includegraphics[height=5.6cm,width=6.8cm]{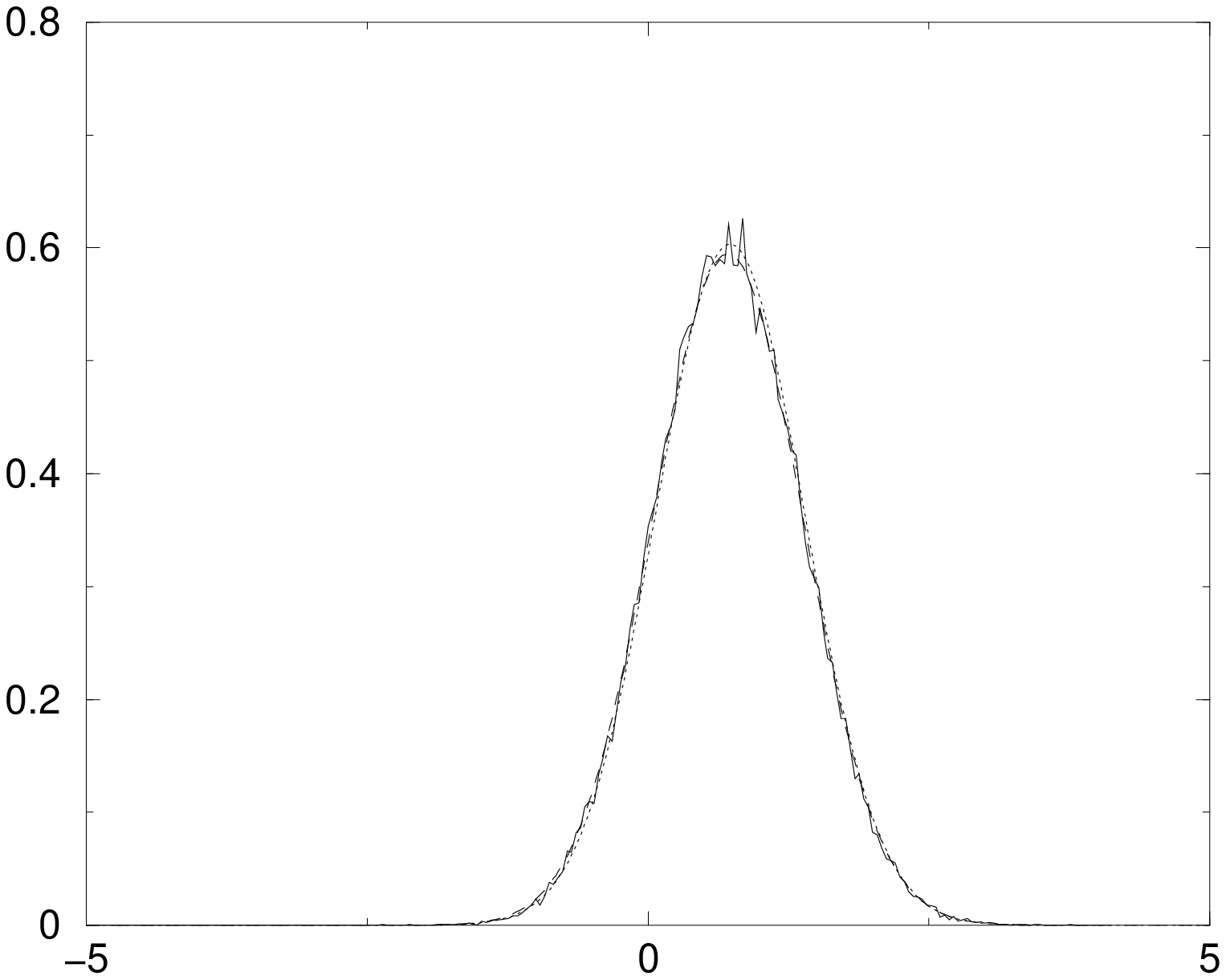}}
\put(90,5){\includegraphics[height=5.6cm,width=6.8cm]{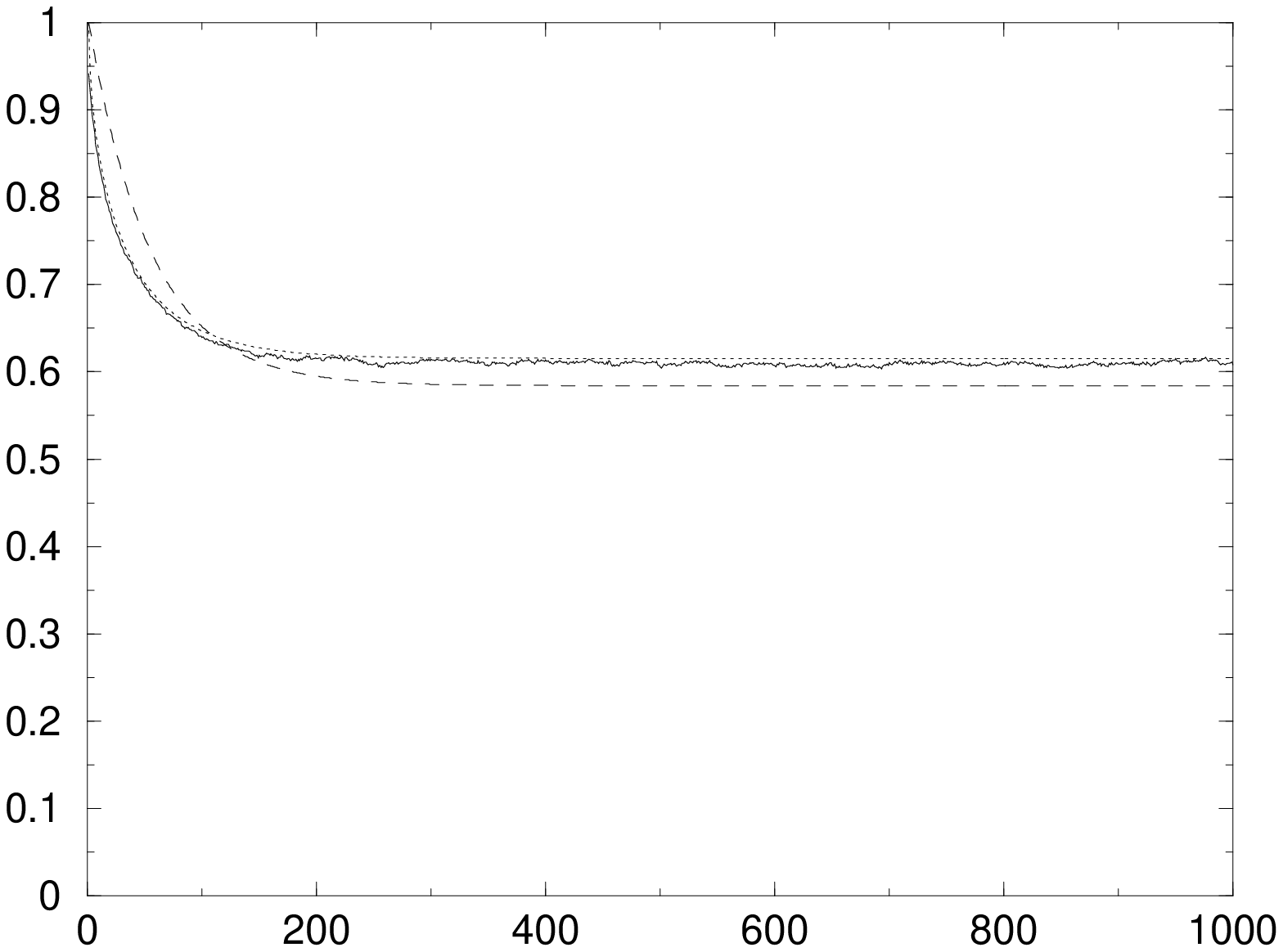}}
\put(44,0){$u$}\put(0,34){$P(u)$} \put(123,0){$t/\Delta
t$}\put(80,34){$C(t)$}
\end{picture}
\caption{\label{CORRDIST1}
         The left graph depicts the static distribution of
         spins at $\alpha = T = 0.25$. The solid line is from a
         simulation of $N=64000$ spins, the dotted line is from solving
         for the correlation function and the dashed line is from the
         interpolation theory. Both theories give excellent agreement
         with the simulation. The right graph depicts the corresponding TTI correlation function as a
         function of the number of discretized time steps $t/\Delta t$ (with here $\Delta t =
         0.02$), with the meaning of the different lines are as in
         the left picture. It is clear that the solving for the correlation
         function numerically gives a significantly better calculation
         of the correlation function.}
\end{figure}

\begin{figure}[t]
\vspace*{3mm} \setlength{\unitlength}{0.1cm}
\begin{picture}(60,60)
\put(10,5){\includegraphics[height=5.6cm,width=6.8cm]{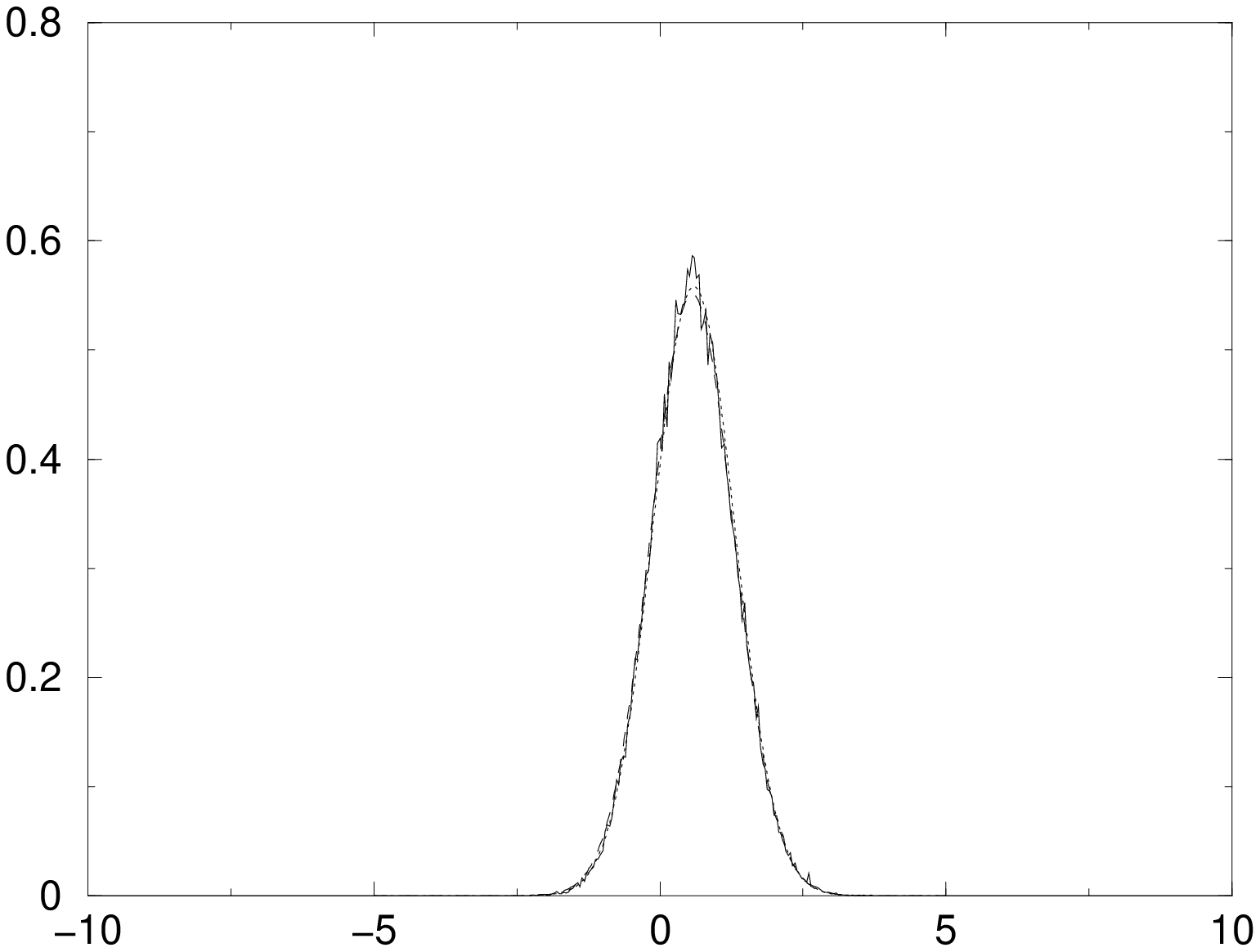}}
\put(90,5){\includegraphics[height=5.6cm,width=6.8cm]{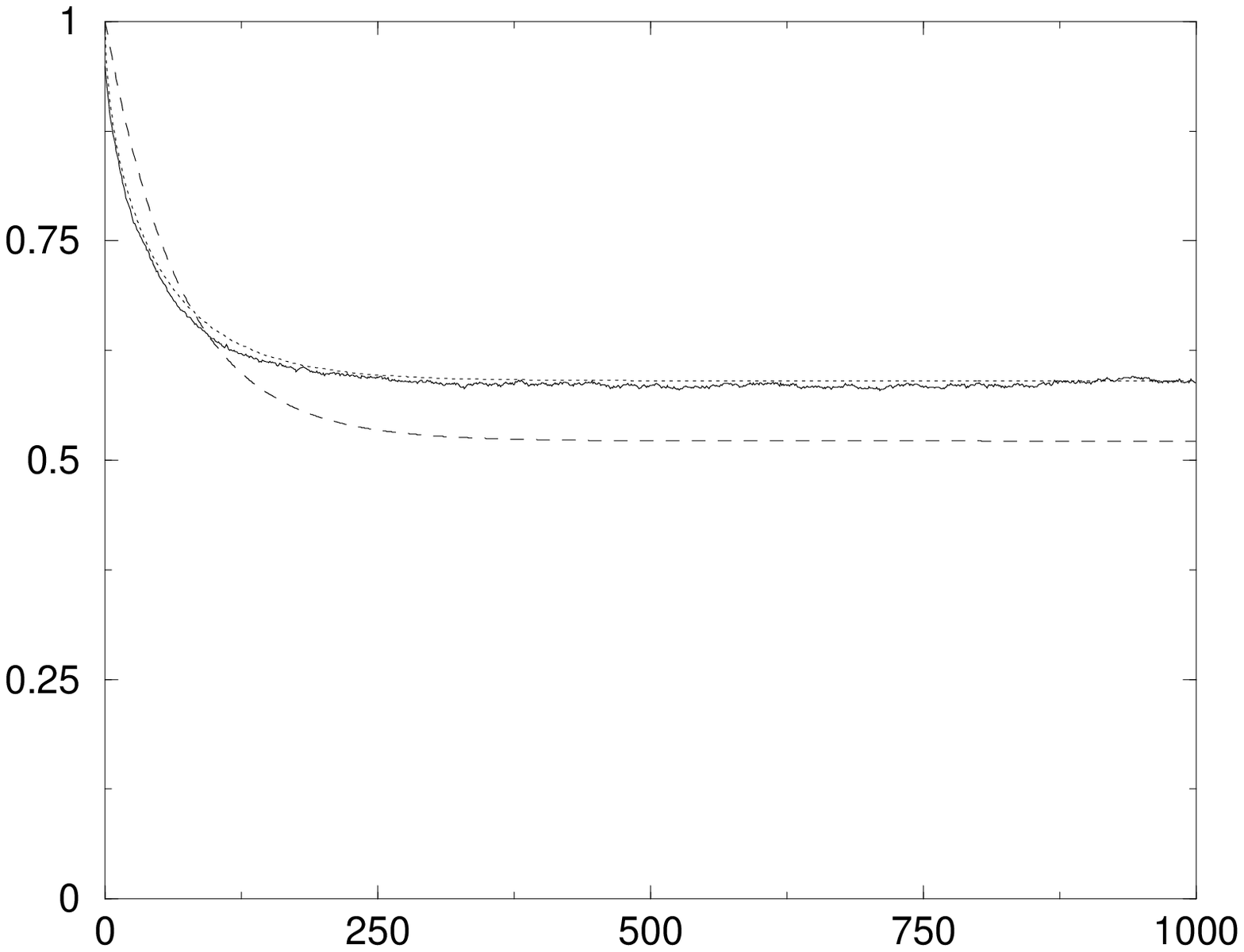}}
\put(44,0){$u$}\put(0,34){$P(u)$} \put(123,0){$t/\Delta
t$}\put(80,34){$C(t)$}
\end{picture}
\caption{\label{CORRDIST2}
         The left graph depicts the static distribution of
         spins at $\alpha=0.5$ and $T = 0.125$. The solid line is from a
         simulation of $N=64000$ spins, the dotted line is from solving
         for the correlation function and the dashed line is from the
         interpolation theory. Both theories give excellent agreement
         with the simulation. The right graph depicts the corresponding TTI correlation function as a
         function of the number of discretized time steps $t/\Delta t$ (with here $\Delta t =
         0.02$), with the meaning of the different lines are as in
         the left picture. It is clear that the solving for the correlation
         function numerically gives a significantly better calculation
         of the correlation function.}
\end{figure}

\section{Comparison with simulations}

In Figures \ref{CORRDIST1} and \ref{CORRDIST2} we examine both the
distribution of spins at a single time and the correlation
function. In simulations the spins evolved for $\mathcal{O}(10^5)$
discretized  time steps (with $\Delta t=0.02$)  in order that they
reached their static distribution. It is clear from
(\ref{eq:usoln}) that the static distribution of spins must be
Gaussian with mean $m$ and variance $\kappa$. Our two theories
(exact versus interpolation) give different values for $m$ and
$\kappa$, yet both are seen to  agree very well with the
simulations as far as the spin distribution is concerned. For the
correlation function, however, it appears that solving the full
order parameter equations numerically  gives excellent agreement
with simulations and is a significantly better guide to both the
persistent correlation and the short time shape than
interpolation. The differences between the two theories appear to
increase with $\alpha$ within the recall regime.

\begin{figure}[t]
\vspace*{3mm} \setlength{\unitlength}{0.1cm}
\begin{picture}(60,60)
\put(10,5){\includegraphics[height=5.6cm,width=6.8cm]{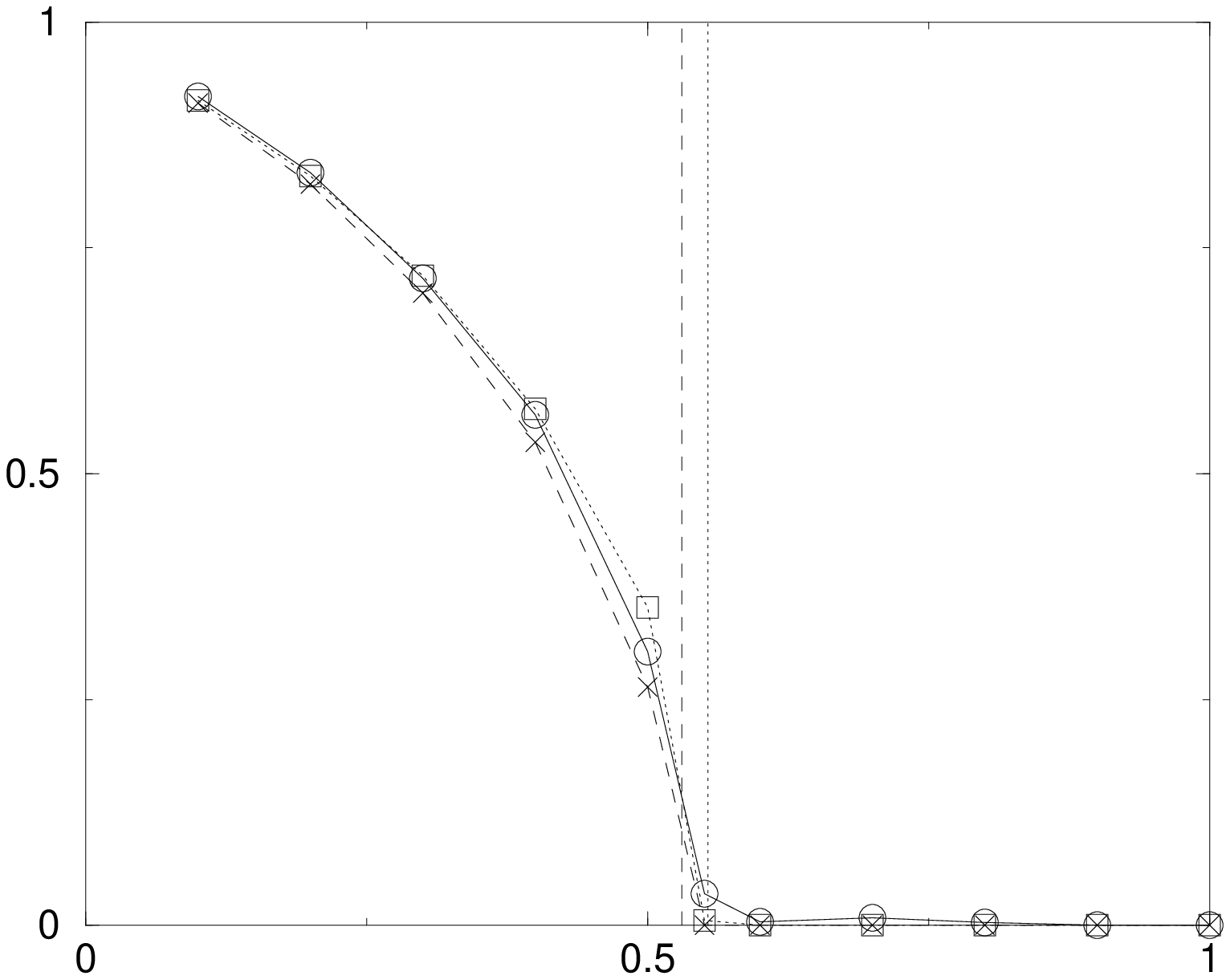}}
\put(90,5){\includegraphics[height=5.6cm,width=6.8cm]{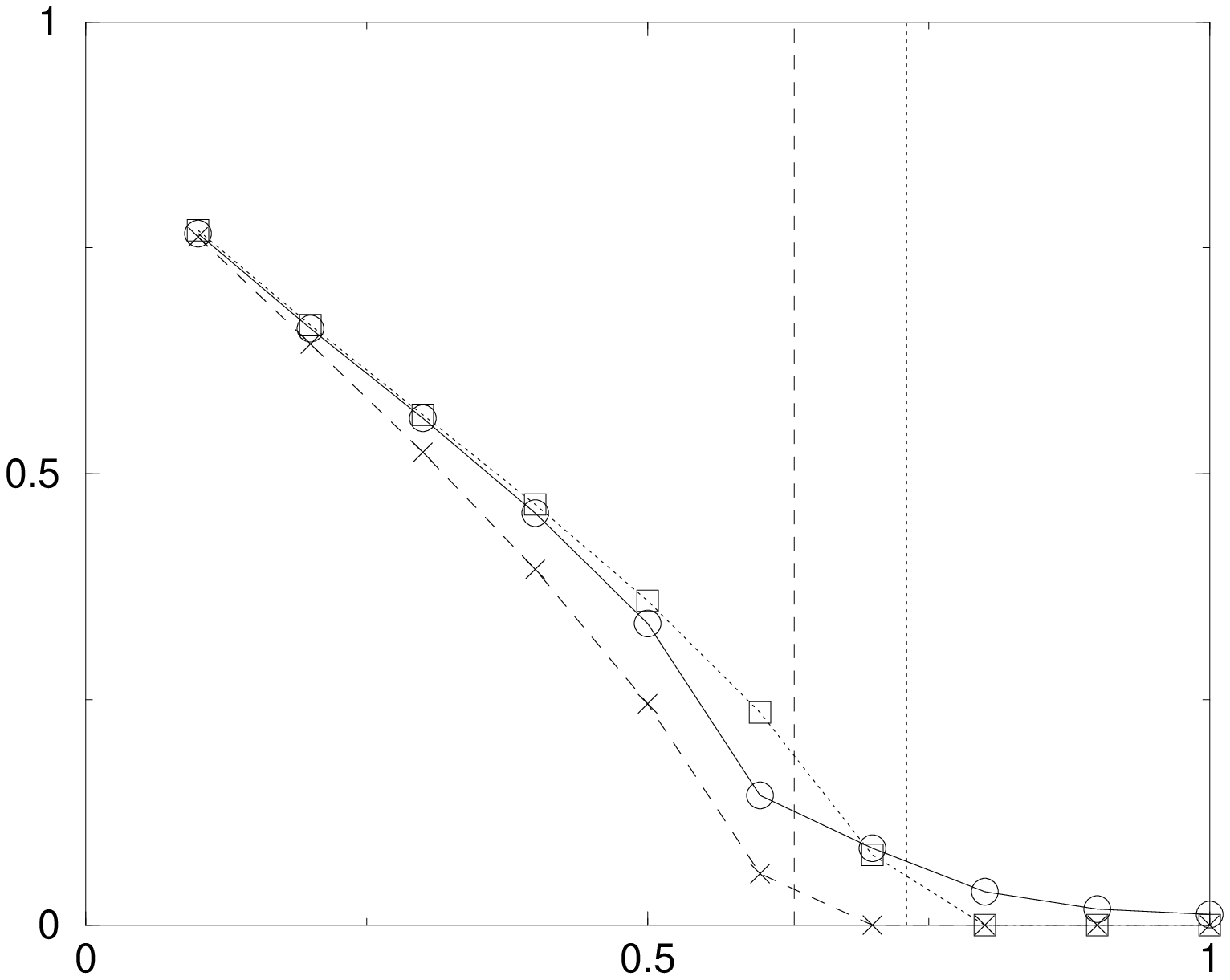}}
\put(43,0){$T$}\put(3,34){$m$}
\put(123,0){$\alpha$}\put(85,34){$q$}
\end{picture}
\caption{\label{Sims}
         The left graph shows the magnetisation versus
         temperature for $\alpha = 0.2$. The solid line with $\circ$ markers is
         averaged
         over 5 simulations of $N=64000$ spins (error bars are all $< 0.01$),
         the dotted line with $\Box$ markers is obtained via solving the correlation
         function
         numerically, and the dashed line with $\times$ markers is calculated from the
         interpolation theory. The vertical lines mark the
         predicted
         phase transitions, the left line being from the
         interpolation.  Both theories give excellent agreement
         with the simulation. The right graph shows the persistent correlation versus alpha at
         $T = 0.25$. The different lines are as
         above. Solving the correlation
         function numerically gives better agreement with the
         simulations at higher $\alpha$.}
\end{figure}

In Figure \ref{Sims} we plot magnetisations $m$ and persistent
correlations $q$ for a variety of values for $T$ and $\alpha$ on
both sides of the predicted phase boundary, for both simulations
and theory. In the left picture we see that for moderate $\alpha
(=0.2)$ both theories give good agreement with each other and with
simulations. However, in the right picture,  i.e. at higher
$\alpha$ near the phase transition, again the numerical solution
of the full order parameter equations gives better agreement.

\section{Discussion}

In this paper we have used the formalism of generating functional
theory to solve the dynamics of an extremely dilute asymmetrically
structured Hopfield-type attractor neural network model with
Langevin dynamics near saturation. This is a non-detailed-balance
problem. In contrast to extremely diluted models with synchronous
dynamics, we here cannot benefit from the Gaussian fields since we
need more than just the marginal field distribution. We have
looked for time translationally invariant solutions as an aid to
finding stationary solutions to this problem. In general we find
that we need a full functional order parameter $C(\tau)$ to
describe this stationary solution exactly. We have also looked at
an interpolative theory that, while only exact at $\alpha = 0$,
depends on only a few scalar parameters rather than an entire
order parameter function. This reduced theory gives good agreement
at low $\alpha$ but shows discrepancies at higher values of
$\alpha$.

As expected for this type of architecture the phase diagrams
exhibit just two phases, a recall phase and a non-recall phase.
There are no spin glass phases, which is due to the lack of a
retarded self-interaction in the effective single spin equation.
We have calculated, for specific choices of the gain function $g$,
the system's $(\alpha, T)$ phase diagram, and have found the
second order transitions in this diagram via a bifurcation
analysis. The methods used in this paper to solve the effective
single spin equation could be valid for other problems with a
similar architecture and similar dynamics. The network's
connection asymmetry is obviously an assumption whose
justification is model dependent, here it is reasonable and leads
to exactly solvable statics.

It would also be of interest to study the initial dynamics of this
problem, as that will be of more relevance to biological
experiments. Other generalizations could include moving to a
finitely connected architecture which may be closer to that seen
in nature. Here the order parameters would become single spin path
probabilities, and the effect on these path probabilities of a
change in the external field at each point on the path
\cite{Guzai2}. It may also be of interest to see if an approximate
theory could be constructed when arbitrary degrees of symmetry are
introduced into the problem, such that  the method presently  used
in this paper is no longer accessible.

\end{document}